\documentclass[conference]{IEEEtran}
\IEEEoverridecommandlockouts
\usepackage{cite}
\usepackage{amsmath,amssymb,amsfonts}
\usepackage{algorithmic}
\usepackage{graphicx}
\usepackage{textcomp}
\usepackage{xcolor}
\usepackage{multirow}

\DeclareMathOperator\sg{sg}
\DeclareMathOperator\CE{CE}
\DeclareMathOperator\MAER{MAER}
\def\BibTeX{{\rm B\kern-.05em{\sc i\kern-.025em b}\kern-.08em
    T\kern-.1667em\lower.7ex\hbox{E}\kern-.125emX}}
\begin{document}

\title{Hierarchical disentangled representation learning for singing voice conversion}

\author{\IEEEauthorblockN{Naoya Takahashi}
\IEEEauthorblockA{\textit{R\&D Center} \\
\textit{Sony}\\
Tokyo, Japan \\
Naoya.Takahashi@sony.com}
\and
\IEEEauthorblockN{Mayank Kumar Singh}
\IEEEauthorblockA{\textit{R\&D Center} \\
\textit{Sony}\\
Tokyo, Japan \\
Mayank.Singh@sony.com}
\and
\IEEEauthorblockN{Yuki Mitsufuji}
\IEEEauthorblockA{\textit{R\&D Center} \\
\textit{Sony}\\
Tokyo, Japan \\
Yuhki.Mitsufuji@sony.com}
}


\maketitle

\begin{abstract}
Conventional singing voice conversion (SVC) methods often suffer from operating in high-resolution audio owing to a high dimensionality of data. In this paper, we propose a hierarchical representation learning that enables the learning of disentangled representations with multiple resolutions independently.  With the learned disentangled representations, the proposed method progressively performs SVC from low to high resolutions.
Experimental results show that the proposed method outperforms baselines that operate with a single resolution in terms of mean opinion score (MOS), similarity score, and pitch accuracy.

\end{abstract}

\begin{IEEEkeywords}
Singing voice conversion, VQVAE, high-resolution
\end{IEEEkeywords}

\section{Introduction}
Singing voice conversion (SVC) is an audio synthesis task to convert some aspects of a singing voice, such as the identity of the singer and pitch, while retaining the musical content such as melody or lyrics. SVC has been attracting increasing attention recently because it may potentially have tremendous impact on a music production community, e.g., singers can improve vocal quality, apply mimicry effects, and even create a virtual singer. 
To tackle this very challenging problem, early SVC approaches relied on parallel data of different singers singing the same song \cite{Kobayashi14,Villavicencio10,Kobayashi15}. Recently, voice conversion approaches for speech that can work with nonparallel data have been proposed \cite{Hsu17,Oord17,Saito18,Kameoka18,Kaneko18,Kaneko19} and successfully applied to SVC, showing promising results \cite{Nachmani19, Chen19, Deng20, Polyak20, Luo20}. 
The basic idea of the nonparallel data-driven SVC approaches is to learn disentangled representations that separately characterize different aspects of singing voice, such as voice color, pitch, lyric, and other musical expressions, from unlabeled data (except singer identity). By manipulating some of the disentangled representations, a jointly learned decoder synthesizes a singing voice of which only some aspects are converted.

Most SVC approaches operate with a low sampling rate, typically 16kHz, to mitigate the difficulty in learning appropriate disentangled representations from high dimensional data. However, professional music contents are usually recorded at a significantly higher sampling rate such as 44.1k or 48kHz. Downsampling the music content to such a low sampling rate considerably degrades the quality of the singing voice. Hence, disentangled representation learning for high-sampling-rate audio is clearly important.

In this paper, we propose a hierarchical disentangled representation learning based on the vector quantized variational autoencoder (VQVAE) to efficiently model high-dimensional audio.  The proposed method learns the disentangled representations with different resolutions independently, which enables the training of an encoder and a decoder with different resolutions in parallel whilst ensuring the quality of SVC in a low sampling rate. 

The contributions of this paper are as follows. (i) We propose a hierarchical vector quantized variational autoencoder (HVQVAE) to improve the quality of SVC for a high sampling rate. To the best of our knowledge, this is the first work to address SVC for a high sampling rate using nonparallel data. (ii) We demonstrate the effectiveness of the proposed method by showing the superior performance over a single-scale method. (iii) We investigate the learned representation to provide insight and show the flexible controllablility of the proposed method. 

\section{Related works}
In this section, we summarize related works on SVC and hierarchical representation learning.
\subsection{Singing voice conversion}
Compared with voice conversion in the speech domain, SVC is more challenging because it needs to deal with a wider range of frequency variation and musical expressions such as vibrato. 
Initial stage of SVC relied on parallel datasets composed of paired samples of different singers singing the same song \cite{Kobayashi14,Villavicencio10,Kobayashi15}. To overcome the limitation of hard-to-obtain parallel data, various approaches that do not require parallel data have been studied recently. In these approaches, a source singing voice is first encoded to disentangled representations by using automatic speech (phone) recognition \cite{Chen19, Polyak20}, a pitch extractor \cite{Chen19,Deng20,Polyak20} or a jointly trained encoder and axial classifier \cite{Nachmani19,Deng20,Luo20}. The disentangled representations are then used to synthesize a singing voice with a vocoder \cite{Chen19} or generative models, such as an autoregressive model \cite{Nachmani19,Deng20}, a variational autoencoder \cite{Luo20} or generative adversarial networks (GANs) \cite{Polyak20}. 
Different from previous works in SVC, we adopt VQVAE \cite{Oord17}, which utilizes a strong information bottleneck effect achieved by discretization, to learn disentangled representations.

\subsection{Hierarchical representation learning}
Hierarchical representation learning for generative models has been investigated in the image domain. In \cite{Razavi19}, discrete representations of an image in multiple resolutions are jointly learned and an autoregressive decoder is successfully learned to generate high-quality high-resolution images. In a hierarchical quantized autoencoder (HQA) \cite{Williams20}, discrete representations are learned hierarchically to achieve an extensive compression rate. The discrete representations with a low resolution are decoded to match representations with a high resolution and again quantized with a stochastic assignment.

In the speech domain, hierarchical modeling methods based on VAE are proposed to provide interpretability and controllability to speech synthesis systems \cite{Sun20,Hsu19}.
In contrast, where representations with multiple resolutions are learned jointly, our method learns representations with different resolutions independently, which enables us to fully enjoy the advantage of low-resolution models and to adaptively change the input resolution in accordance with the computational resource or bandwidth requirements.

Apart from autoregressive models, PG-GAN \cite{Karras18} can be also considered as a way of learning representations in a hierarchical manner. At the beginning of PG-GAN training, a generator learns to generate low-resolution images, and the generator is progressively extended to generate high-resolution images.

\section{Method}
We first summarize the VQVAE-based representation learning. Then, we adopt it to SVC. Finally, the hierarchical representation learning method is introduced.  

\subsection{Single scale VQVAE}
VQVAE is an extension of VAEs that include discrete latent variables. It consists of an encoder $E(x)$ that maps input $x$ to a latent space, followed by a vector quantization operation $V(E(x)) = z_q$ that assigns the nearest embedding $e_k\in\mathbb{R}^{D_e}$ from a codebook $\mathcal{C}=\{e_k|k=1,\dots,K\}$, and a decoder with a distribution $p(x|z_q)$ that recovers the input signal. The encoder, decoder, and embeddings in the codebook are jointly trained by maximizing the evidence lower bound (ELBO) while regularizing the encoder to commit an embedding and updating the embeddings to a mean of assigned encoder outputs. By defining a uniform prior over $z$, the loss function $L$ results in
\begin{equation}
\label{eq:vqvae}
L= - \log p(x|z_q) + \|\sg(E(x))-e\|^2 + \beta\|E(x) - \sg(e) \|^2,
\end{equation}
where $\sg()$ is the stopping gradient operation to exclude the argument from the gradient calculation.
The first term corresponds to the negative ELBO and is called the reconstruction loss $L_{rec}$ (the KL term of ELBO becomes constant owing to the uniform prior over $z$, and hence, it is omitted). The second term is the objective for updating the embeddings, and the last term with weight $\beta$ is the commitment loss for regularizing the encoder.
Owing to the discrete latent representation and the strong information bottleneck induced by vector quantization, VQVAE demonstrates an excellent capability of capturing a discrete nature of data such as phones in speech and object categories in images \cite{Oord17}.

\begin{figure}[t]
  \centering
  \includegraphics[width=\linewidth]{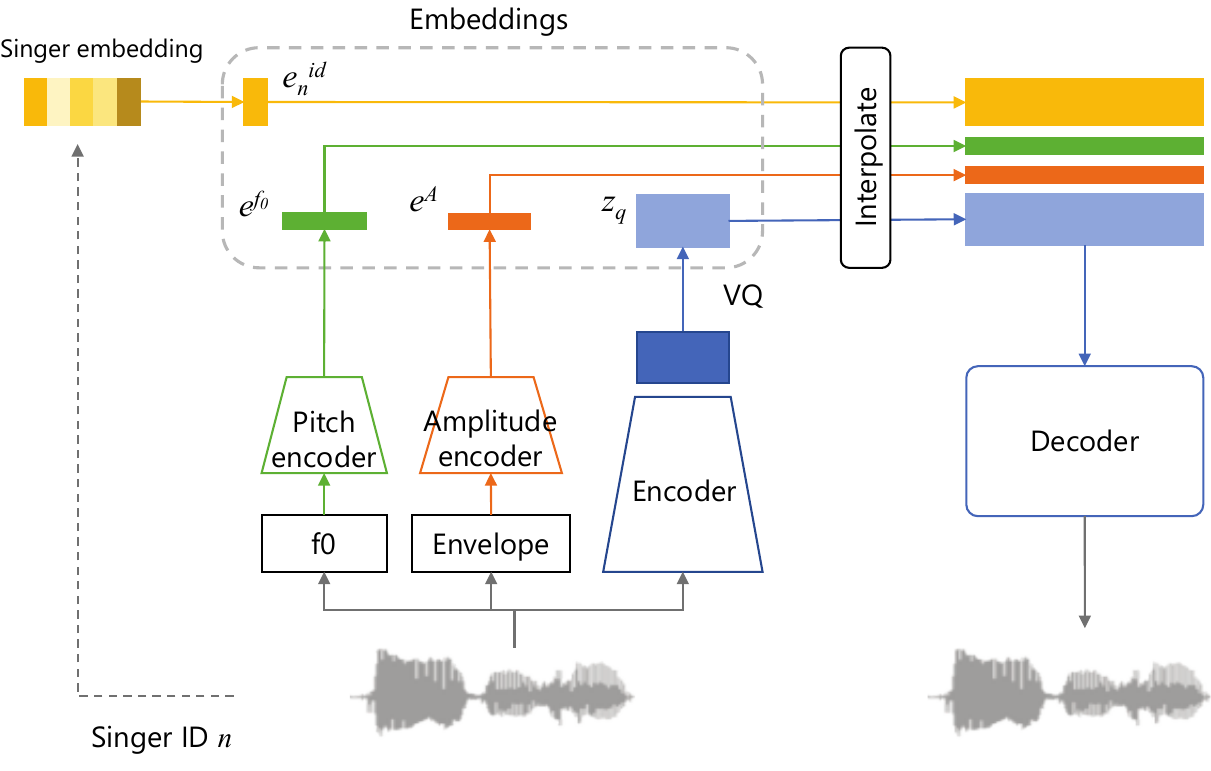}
  \caption{VQVAE-based SVC. Singer embedding $e^{id}$, encoders, and a decoder are jointly trained. }
  \label{fig:VQVAE_SVC}
\end{figure}

\subsection{VQVAE-based SVC}
\label{sec:VQVAE-SVC}
Simply training  VQVAE on a singing voice dataset with \eqref{eq:vqvae} may not sufficiently disentangle some aspects that we want to control, such as singer identity, pitch, and loudness. To ensure the controllability of these aspects that are required for SVC, in addition to $z_q$, we design the model to learn embeddings from extracted features that correspond to the aspects, as shown in Fig. \ref{fig:VQVAE_SVC}. Specifically, we extract fundamental frequency $f_0\in \mathbb{R}^{T_f}$ and envelope $A\in\mathbb{R}^{T_A}$ from an input wavefrom and map them to pitch embeddings $e^{f_0}\in \mathbb{R}^{D_f\times T_f}$ and loudness embeddings $e^{A}\in\mathbb{R}^{D_A\times T_A}$ using feature encoders, where ${T_f}$ and ${T_A}$ are the time steps and ${D_f}$ and ${D_A}$ are the dimensions of the embedding vectors. The model also jointly learns time-invariant singer embedding $e_n^{id}\in\mathbb{R}^{D_{id}}$ for each singer. All embeddings are used to condition the decoder as $p(x|z_q, e^{f_0}, e^A, e_n^{id})$. Explicitly including the feature embeddings $e^{f_0}, e^A, e_n^{id}$ promotes $z_q$ to be independent of the pitch, loudness, and singer identity. For clarity, we call $z_q$ a content embedding.
As we use the WaveNet \cite{Aaron2016WN} decoder, a powerful autoregressive model, all embeddings are  linearly interpolated to have the same number of time steps, $T$, as input $x\in\mathbb{R}^T$. 

During training, the singer embedding corresponding to the input (source singer) is fed to the decoder and embeddings are jointly trained with the encoder and decoder using \eqref{eq:vqvae}. At the inference time, the singing voice can be converted by manipulating embeddings, e.g. by providing a target singer embedding different from the source singer embedding to convert the singer identity, or by changing $f_0$ to convert the pitch.

Note that there is no guarantee that the content embeddings $z_q$ do not contain any information about the fundamental frequency, loudness, or singer identity as we do not explicitly remove such information from it. If the model fails to disentangle such aspects and the content embeddings $z_q$ retain such information, the decoder may neglect $e^{f_0}$ or $e_n^{id}$, and SVC with their manipulation may not happen.
Nevertheless, as we show in our experiments, the strong information bottleneck induced by the quantization of the encoded feature tends to remove such information and the decoder relies on $e^{f_0}$ and $e_n^{id}$ to synthesize the singing voice, enabling SVC. One can also consider including adversarial loss on the content embeddings; however, we focus on investigating the effect of vector quantization.

\subsection{Hierarchical VQVAE (HVQVAE)}
\begin{figure}[t]
  \centering
  \includegraphics[width=\linewidth]{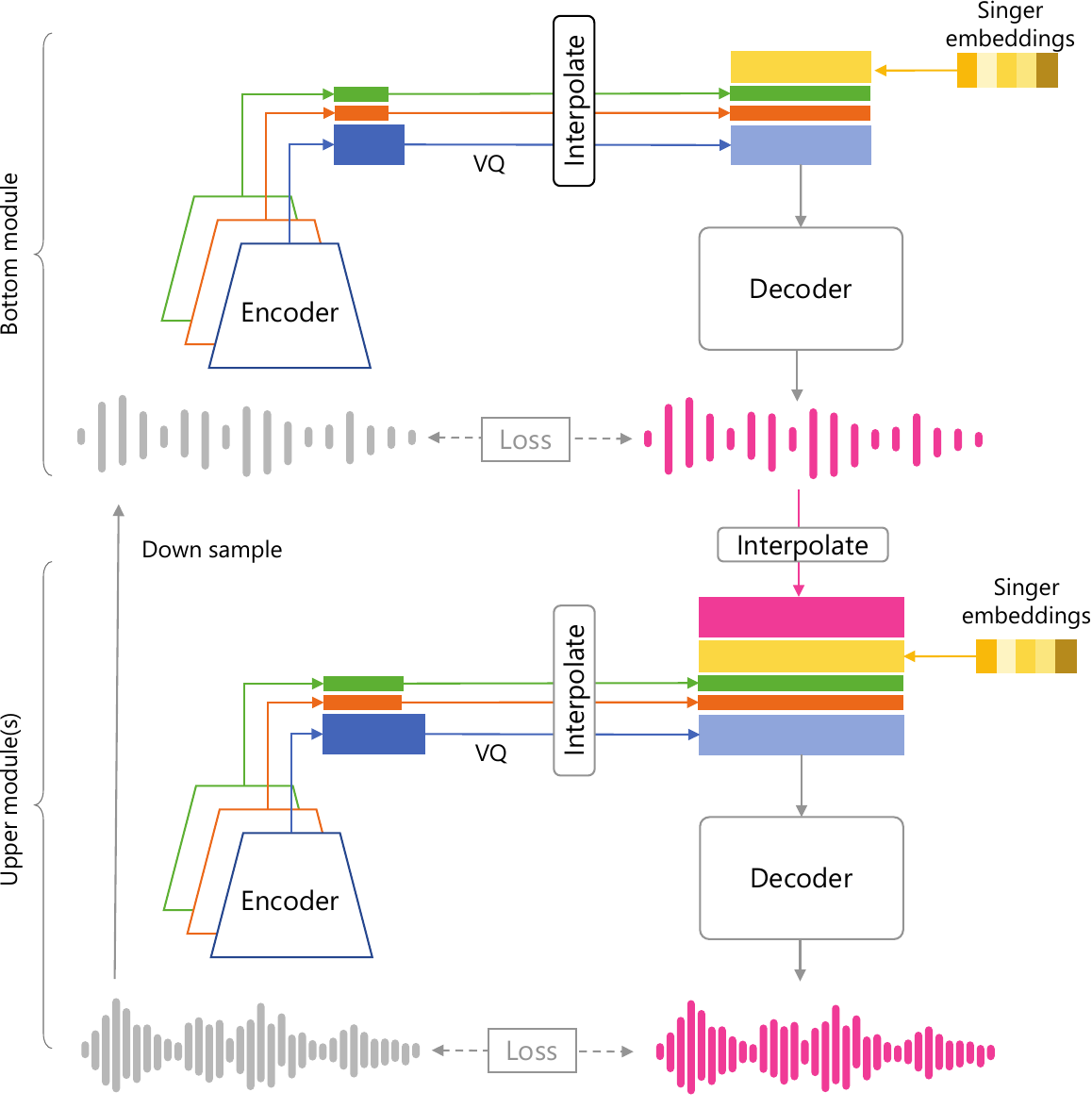}
  \caption{Illustration of HVQVAE. The bottom module works on dowsampled data. The upper modules work on higher-sampling-rate input data and the decoder is conditioned on the output of the lower module.}
  \label{fig:HVQVAE}
\end{figure}
As the sampling rate increases, the model must learn to encode higher-dimensional input to latent disentangled representations and to synthesize higher-dimensional data to produce a same-length audio, which makes the task increasingly difficult. To overcome this problem, we propose a hierarchical representation learning, HVQVAE as shown in  Fig. \ref{fig:HVQVAE}. HVQVAE consists of a bottom VQVAE module and one or more upper VQVAE modules. The bottom VQVAE module is trained and performs exactly the same as the single-scale VQVAE, as discussed in Sec. \ref{sec:VQVAE-SVC}, except that the input is downsampled to a lower resolution to make the representation learning easy. By operating on low-resolution input, the bottom module can more easily capture a global structure (low-frequency components) of data.
The upper module utilizes the output of the lower module to condition the decoder in addition to other embeddings encoded from the higher-resolution input. Since the global structure of data is already modeled in the lower module and the low-resolution output is provided, the upper module can focus on learning representations for detailed structures (high-frequency components), which facilitates the learning of the representation for high-resolution data. Multiple upper modules can be stacked hierarchically to gradually decrease and increase the resolution of data. 
During the training of the upper module, we use downsampled  input data rather than the reconstruction of the lower module to condition the decoder of the upper module. This allows us to train each module independently. Note that owing to the nature of the autoregressive decoder, jointly training all modules using the outputs of the lower module is not feasible and may even not be optimal.
Another advantage of the proposed hierarchical representation learning is that outputs from all intermediate decoders are ensured to be audio with different sampling rates, rather than latent representations. This enables us to process audio with different sampling rates in accordance with the requirements of the computational resource or bandwidth in an ad-hoc manner.

\begin{figure}[t]
  \centering
  \includegraphics[width=0.6\linewidth]{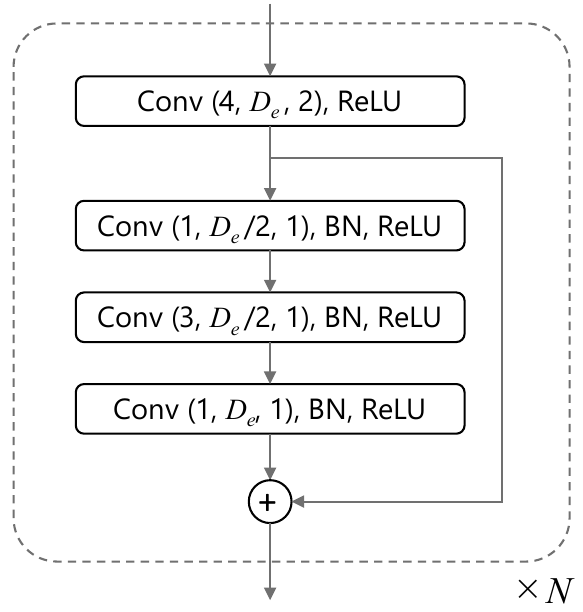}
  \caption{The convolution block used in the encoder. Each convolution layer is specified by Conv$(W, C, S)$, where $W$ denotes the kernel size, $C$ the number of output channels, and $S$ the stride. All convolutions and BN are one-dimensional.}
  \label{fig:convblock}
\end{figure}

\subsection{Implementation detail}
In the rest of our paper, we consider the case where the number of upper modules is one, i.e., two-scale HVQVAE, and the bottom module performs with a six-times-lower resolution than the upper module. 
The encoder $E(x)$ consists of convolution blocks as illustrated in Fig. \ref{fig:convblock}. The convolution block first downsizes the feature map by a factor of two with a stride convolution, followed by a residual block with a bottleneck structure. The convolution blocks are stacked $N$ times; therefore, the embeddings are obtained with a $2^N$-times-lower resolution. We use $N=6$ for the bottom VQVAE module and $N=8$ for the upper module. The feature encoders for $f_0$ and the envelope consist of three 1D convolution layers with dilations of 1, 2, and 4, each followed by ReLU nonlinearity. The dimensions of embeddings $D_e$, $D_{f_0}$, $D_{A}$, and $D_{id}$ are 512, 10, 10, and 128, respectively. The size of the discrete embedding space (codebook size) $K$ is 320. The decoder is WaveNet \cite{Aaron2016WN} with the channel size of 64 and 30 layers. Audio data are transformed with $\mu$-law encode at the 256 quantization level, and the decoder outputs the posterior over the 256-way categorical distribution. Thus, the reconstruction loss term in \eqref{eq:vqvae} becomes the cross-entropy,

\begin{equation}
\label{eq:hvqvae}
L_{rec}=\sum_{t=1}^{T-1}\CE(p(x_t|x_{t-1,\dots,0}, z_{q}, x^{low}, e^{f_0}, e^A, e_{n}^{id}),x_t),
\end{equation}
where $\CE(p,y)$ is the cross-entropy loss, $p$ is the predicted distribution, $y$ is the label, $x_t$ is the value of the signal at time step $t$, and $x^{low}$ is the low-resolution input.

\section{Experiments}

\subsection{Dataset}
We use the NUS-48E dataset \cite{Duan13} for experiments. The dataset contains 48 songs sung by 12 singers (6 male and 6 female), 4 songs for each singer. In parallel, the dataset also contains the ``\textit{read}'' style of these songs, i.e., singers read the lyrics of the same song with neither musical melody nor rhythm. We use both styles for training. Audio files are provided at a sampling rate of 44.1kHz in stereo or mono format. We convert them to the 48kHz, mono format.

\subsection{Setup}
We randomly sample short frames from both \textit{sing} and \textit{read} data during the training. The frame length is 7680 samples and the batch size is 32. 
The regularization weight $\beta$ in \eqref{eq:vqvae} is set to 0.25. The model is trained for 20K iterations using the Adam optimizer with the learning rate of 0.0002.
The WORLD vocoder \cite{WORLD} is used to extract the fundamental frequency with 5 ms period. For envelope extraction, we use the average absolute amplitude of the waveform in the window size of 40 ms with half-overlap (20 ms shift).

\subsection{Baseline}
To evaluate the effectiveness of the hierarchical representation learning, we compare the proposed method with a standard VQVAE, which directly learns representation in a single scale from the high-resolution data. For fair comparison, we design VQVAE to have similar information compression rate for the embedding space. To this end, we set $N=8$ to learn embeddings in the same scale as HVQVAE and increase the codebook size to $K=640$, since HVQVAE has two modules with $K=320$ each. Other parameters are exactly the same as those of HVQVAE.

\subsection{Singer identity conversion}
We first evaluate the quality of singer identity conversion though a subjective test. We use two metrics, mean opinion score (MOS) and similarity score. For MOS, raters are asked to rate the naturalness of a converted singing voice from 1 (totally unnatural) to 5 (completely natural). For similarity score, raters listen a pair of an unprocessed reference voice and the converted sample, and rate the similarity from 1 (dissimilar) to 5 (very similar). We ensure that the reference and sample are from different songs. As a reference, we also evaluate the ground truth. For each experiment, 20 raters evaluate 12 samples, resulting in 240 evaluations for each model. 

The results are summarized in Table \ref{tab:subj}. VQVAE suffers from modeling the high-resolution audio and synthesizes unnatural converted singing voices.  HVQVAE greatly improves the naturalness, achieving a 1.15 higher MOS than VQVAE. These results show that the hierarchical learning facilitates the modeling of high-resolution audio.  The similarity score of HVQVAE is slightly improved, but is not as significantly as MOS. 

\begin{table}[t]
    \caption{\label{tab:subj} {Test scores for singer identity conversion on NUS-48E dataset.}}
    \centering{
    \begin{tabular}{c | c c } 
    \hline
    Method & MOS $\uparrow$ & Similarity $\uparrow$\\
    \hline\hline
    Ground Truth	& 4.44 $\pm$ 0.27 &	4.28 $\pm$ 0.26  \\
    \hline
    VQVAE (Baseline) &	1.82 $\pm$ 0.14 & 2.35 $\pm$ 0.23 \\
    HVQVAE (Proposed) &	2.97 $\pm$ 0.18 & 2.47 $\pm$ 0.26 \\
    \hline
    \end{tabular}
    }
\end{table}

\subsection{Pitch conversion}
Next, we tested the accuracy of pitch conversion. For this, we extract $f_0$ from input audio and convert it to one semitone ($\simeq$ 6\%) higher or lower to condition the decoder. From the synthesized audio, we again extract $f_0^{rec}$ and calculate the mean absolute error (MAE) relative to the original pitch $f'_0$ used to condition the decoder. We also report the mean absolute error ratio (MAER), defined as

\begin{equation}
\label{eq:maer}
\MAER = \frac{1}{M}\sum_{i=1}^M \frac{|f_0'- f_0^{rec}|}{f'_0},
\end{equation}
where $M$ is the number of test samples.

\begin{figure}[t]
  \centering
  \includegraphics[width=\linewidth]{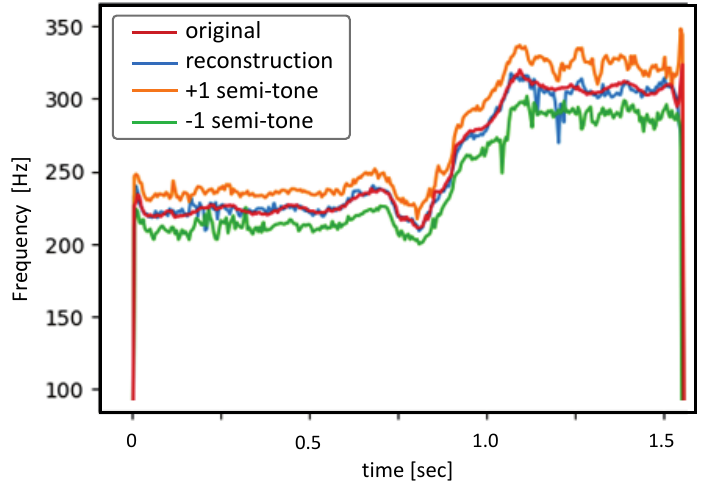}
  \caption{Visualization of f0 sequences extracted from synthesized singing voices. }
  \label{fig:f0}
\end{figure}

The results in Table \ref{tab:pitch} show that HVQVAE successfully converted the pitch with a high accuracy of $1.1\%$ MAER. Compared with the case where pitch is not converted, MAERs of the pitch-converted cases are almost the same, which demonstrates that the decoder solely depends on the pitch embedding to recover the pitch of the synthesized audio, indicating that the pitch embedding is well disentangled.
HVQVAE again shows the superior accuracy of pitch control over the baseline VAVQE model. An example of the pitch sequences is shown in Fig. \ref{fig:f0}.

\begin{table}[t]
    \caption{\label{tab:pitch} {Test scores for pitch conversion. "no shift" refer to the case where the pitch extracted from input is directly used for synthesizing singing voices without modification.}}
    \centering{
    \begin{tabular}{c c | c c } 
    \hline
    Method & Pitch & MAE [Hz] $\downarrow$ & MAER [\%] $\downarrow$\\
    \hline\hline
    \multirow{3}{*}{VQVAE (Baseline)} & no shift & 3.94 &	1.8\\
    & + 1 semitone &	3.80 &	1.5\\
    & - 1 semitone &	3.53 &	1.6\\
    \hline
    \multirow{3}{*}{HVQVAE (Proposed)} & no shift	& 2.41 &	1.0\\
    & + 1 semitone &	2.85 &	1.1\\
    & - 1 semitone &	2.40 &	1.1\\
    \hline

    \end{tabular}
    }
\end{table}

\subsection{Dynamics conversion}
\begin{figure}[t]
  \centering
  \includegraphics[width=\linewidth]{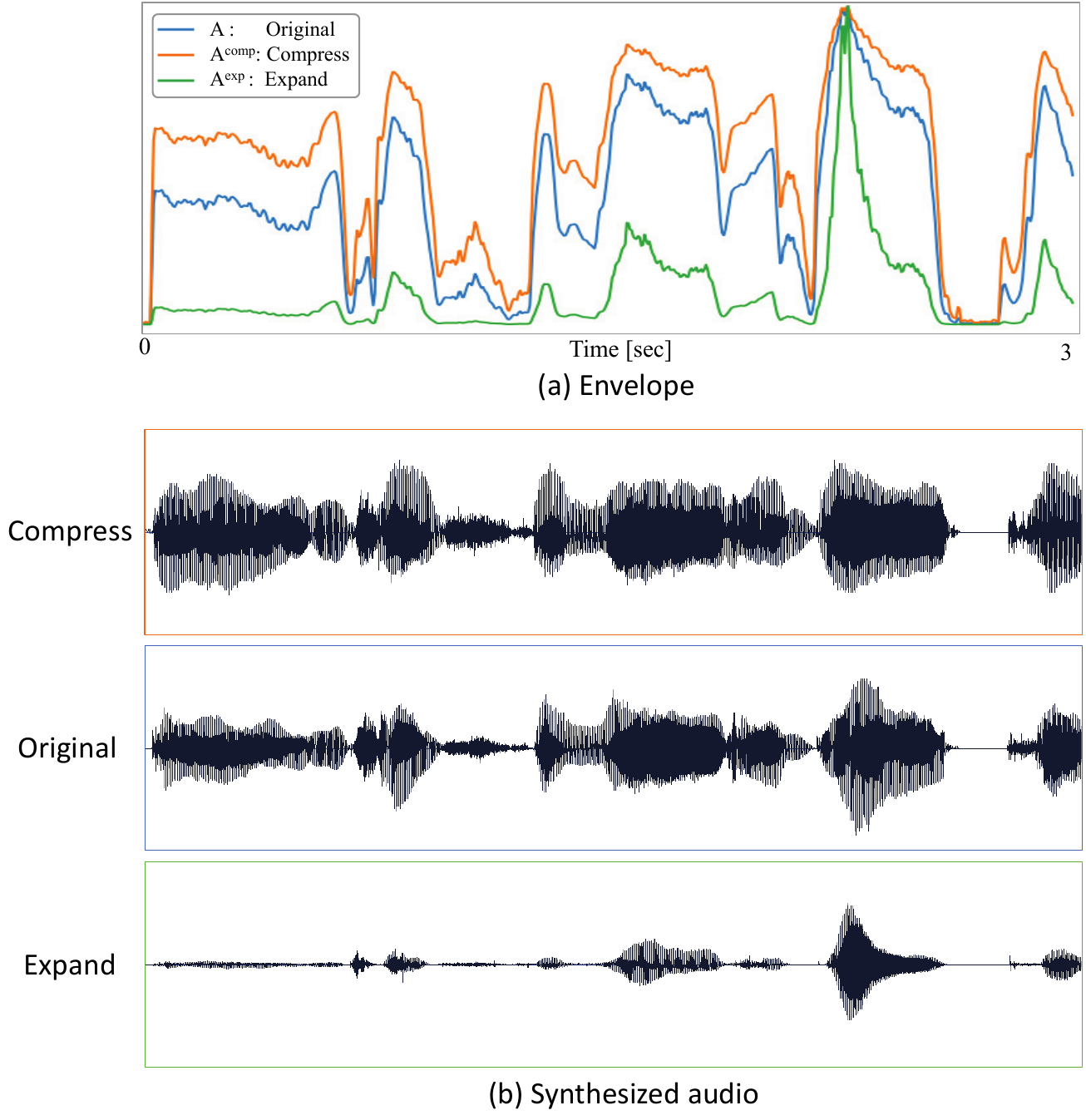}
  \caption{An example of dynamics conversion. (a) Envelopes used for synthesizing audio. (b) Synthesized audio. The proposed method successfully reproduce the provided envelope information.}
  \label{fig:drc}
\end{figure}

As our model disentangles the loudness information, the dynamics of the singing voice is controllable by manipulating the envelop used to extract the loudness embedding $e^A$. We demonstrate the controllability of the dynamics by synthesizing audio using expanded and compressed envelopes. The expanded envelope $A^{\exp}$ is calculated from the original envelope $A$ as
\begin{equation}
\label{eq:exp}
A^{\exp} = A_{max}\frac{e^{\theta\frac{A}{A_{max}}}-1}{e^{\theta}-1},
\end{equation}
and the compressed envelope $A^{comp}$ is obtained  as 
\begin{equation}
\label{eq:exp}
A^{comp} = A_{max}\frac{\log{\theta\frac{A}{A_{max}}}+1}{\log{\theta}+1},
\end{equation}
where $A_{max}$ is the maximum amplitude.
The converted envelope is shown in Fig. \ref{fig:drc}a. The compressed and expanded envelopes have the same peak value as the original envelope, but the dynamic range is compressed towards the maximum amplitude in the compressed envelope, whereas the dynamic range becomes wider in the expanded envelope. Conditioned with loudness embeddings extracted from these envelopes, the HVQVAE decoder successfully synthesizes singing voices that reflect the provided envelope as shown in Fig. \ref{fig:drc}b.    

\subsection{Learned embeddings}
\begin{figure}[t]
  \centering
  \includegraphics[width=\linewidth]{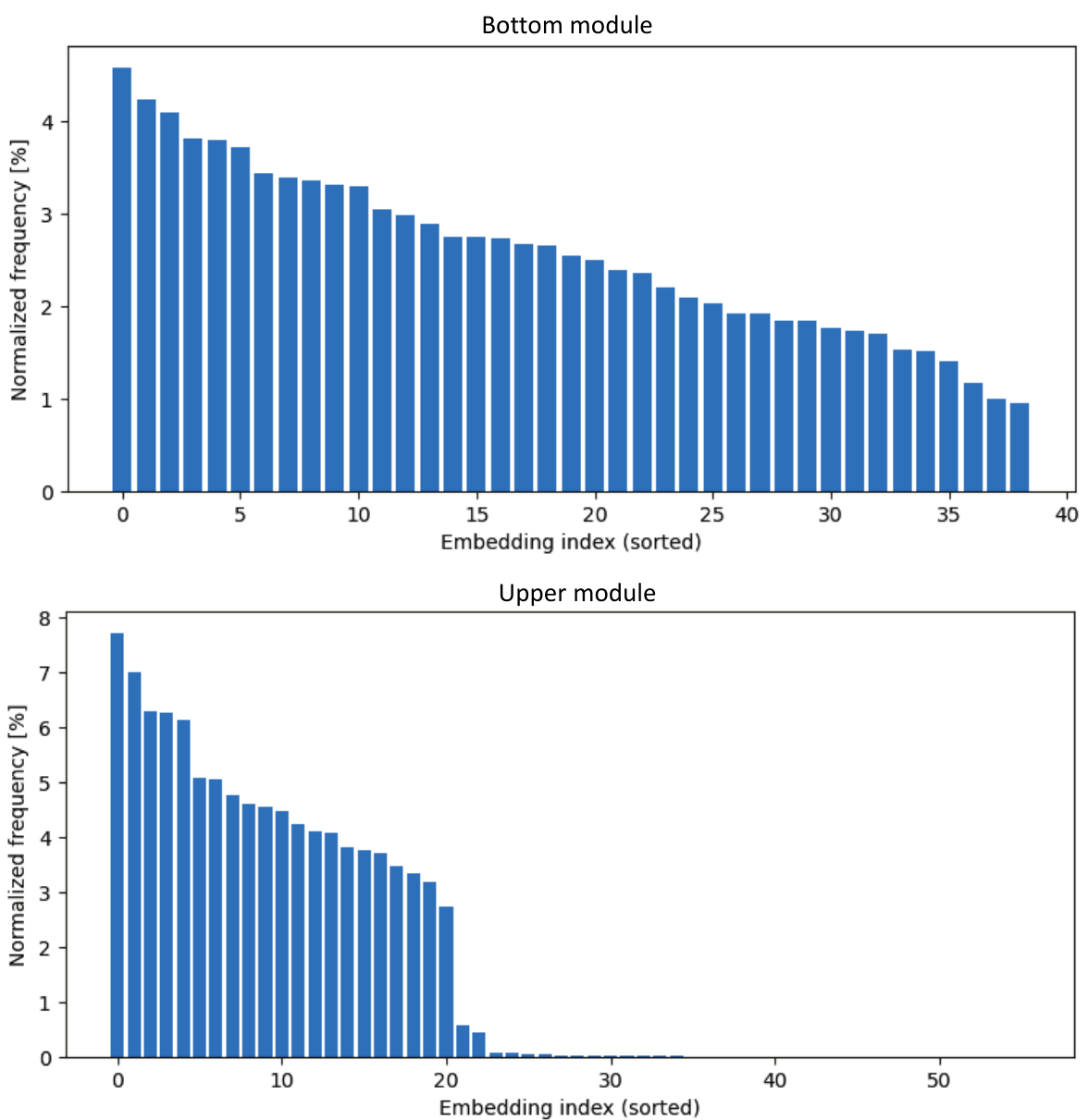}
  \caption{Normalized frequency of embeddings used for encoding NUS48E dataset. Unused embeddings are omitted.}
  \label{fig:histgram}
\end{figure}

We first investigate the learned representations by plotting the frequency of each embedding used for encoding the NUS48E dataset. 
We encode all 48 songs (\textit{``sing"}-style subset) using the learned encoder and determine the frequency of each quantized embedding in the codebook. 
Interestingly, we find that only 39 codes out of $K=320$ are used for encoding all songs in the bottom module, and 56 codes are used in the upper module. These numbers are roughly the same as the number of phones used in the dataset (41). The normalized frequency of embedding usage is shown in Fig. \ref{fig:histgram}, where each bar shows how many times each embedding is used for encoding the dataset. We omit unused embeddings from the plot. In the bottom module, the usage of embeddings is balanced, all embeddings are used from 1 to 4\% of the time. This implies that the model learns to effectively use a small number of embeddings to represent a singing voice. 
In the upper module, on the other hand, only 20 embeddings are frequently used and 36 embeddings are rarely used. We found that these rarely used 36 embeddings are close to each other in the embedding space, which means that a small fluctuation of encoder output in that region can lead to a different assignment of the code by vector quantization and, therefore, the frequency of embedding usage in the region is spread to the 36 embeddings. The number of frequently used embeddings is less than that in the bottom module. This implies that a smaller number of codes are required for characterizing the higher-frequency components (over 8kHz in our case) of singing voices. This is reasonable because some phones do not contain high-frequency components and some phones can be characterized only by low-frequency components.

\begin{figure}[t]
  \centering
  \includegraphics[width=\linewidth]{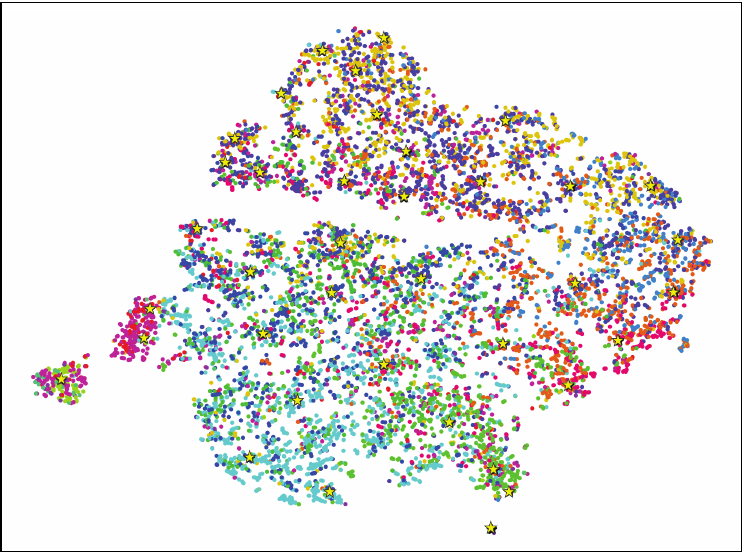}
  \caption{Visualization of embeddings using tSNE. Dots are the encoder outputs before vector quantization. Colors are assigned using the phone label provided in the NUS-48E dataset. Yellow stars are the quantized content embeddings. We observe a weak correlation between phones and embeddings.}
  \label{fig:tSNE}
\end{figure}

We also investigate the relationship between the content embeddings and phones by visualizing the encoder outputs (before vector quantization) and content embeddings using tSNE \cite{tSNE}. We assign colors to encoder outputs on the basis of time-aligned phone-level manual annotation provided in the NUS-48E dataset.  As shown in Fig. \ref{fig:tSNE}, some phones are clustered in a relatively small region, which implies that some phones are weakly correlated with encoder outputs and nearby embeddings. On the other hand, some phones are scattered over a wide area, suggesting that the model learns to encode and discretize singing voices in a different way from linguistic representation. Further assessment on what property is correlated with the learned representation will be a topic of future work. 



\subsection{Ablation study on higher-resolution module}
Since the low-resolution SVC is already achieved in the bottom module, the upper module can be considered as a super-resolution module that is conditioned on embeddings provided by encoders in the upper module. Therefore, some questions naturally arise:``\textit{Does the decoder in the upper module really depend on the embeddings?}", and ``\textit{Are the embeddings encoded in the upper module really necessary to achieve the super-resolution?}" To answer these questions, we consider three baselines for the upper module. (i) To examine the dependence of the decoder on the embeddings, we add a 20\% level of noise to the extracted features $f_0$ and $envelop$, and feed them to the feature encoders. For the quantized content embedding $z_q$, we randomly sample the embeddings from the learned codebook $\mathcal{C}$ by following the same distribution as the provided input. (ii) To test whether the decoder performs singer-dependent super-resolution, we provide the singer embedding of the source singer instead of the target singer. (iii) We train a decoder that solely depends on the output of the bottom module. This setting is considered as the standard super resolution model. Note that in all cases, the bottom module is the same.

We conducted a subjective pairwise comparison test for the evaluation. 
The singer identity is converted by each baseline and proposed method, and synthesized voices of proposed method are compared with those of baselines. Subjects are asked to chose which one of them is preferable in terms of quality.
Each subject evaluates 6 pairs for each baseline. Table \ref{tab:pwcomp} shows the ratio of the proposed method being preferred against the baselines. Clearly, the proposed method is preferred more frequently than all baselines. With noisy embeddings, the upper module produces low-quality voices, showing that the decoder indeed depends on the embeddings. 
When the singer embedding of the source singer is used in the upper module, there is a mismatch with the output of bottom module, which was conditioned on the target singer embedding.
The subjects tend to prefer the result of the proposed method, showing that the upper module indeed performs singer-dependent super-resolution, although the baseline model is still able to produce a reasonable output since pitch-, envelope-, and arguably phone-related information ($e^{f_0}$, $e^A$, $z_q$) provided by the encoders are the correct ones. 
Lastly, the decoder that depends only on the output of the bottom module performs poorly. We found that the baseline model tends to produce a high level of noise around the region where the input has high-frequency components, such as fricatives. This result supports the finding that learning embeddings in the upper module is necessary to synthesize a high-quality singing voice.

\begin{table}[t]
    \caption{\label{tab:pwcomp} {Pairwise comparison of different upper modules. The ratio indicates the frequency of the proposed method being preferred (higher is better).}}
    \centering{
    \begin{tabular}{c | c c} 
    \hline
    Compared upper module & Preferred ratio $\uparrow$ \\
    \hline\hline
    Noisey $f_0$, $envelop$, $z_q$ &	76\% \\
    Source singer ID & 55\%\\
    without embeddings & 100\%\\
    
    \hline
    \end{tabular}
    }
\end{table}

\section{Conclusion}
In this paper, we proposed a hierarchical disentangled representation learning based on VQVAE for singing voice conversion of high-resolution audio. The proposed method consists of multiple VQVAE modules: each module learns the representation with different resolution independently. During the inference time, the module in higher resolution synthesizes a singing voice by using the output of a module with lower resolution. Experimental results on high-resolution singing voice conversion tasks show that the proposed method outperforms a single-scale VQVAE baseline method in terms of naturalness, similarity, and pitch accuracy. The ablation study further confirms the importance of the hierarchical representations.

\bibliographystyle{IEEEtran}
\bibliography{vc}

\end{document}